\documentclass[twocolumn,tighten]{aastex63}

\usepackage{color,amsmath}
\newcommand\Msun{\mbox{$M_\sun$}}
\newcommand\Mspy{\mbox{\Msun\,yr$^{-1}$}}
\newcommand\kmps{\mbox{km\,s$^{-1}$}}
\newcommand\Vexp{\mbox{$V_{\rm exp}$}}
\newcommand\vct[1]{\mbox{$\vec{\mathbf{#1}}$}}%
\newcommand\vcttwo[2]{\mbox{$\vec{\mathbf{#1}}_{\mathbf{#2}}$}}%

\shorttitle{Pinwheel in a triple system}
\shortauthors{Kim et al.}

\received{2023 December 15}
\revised{2024 March 4}
\accepted{2024 March 15}

\begin{document}
\title{Pinwheel Outflow induced by Stellar Mass Loss in a Coplanar Triple System}

\author[0000-0001-9639-0354]{Hyosun Kim}
\affiliation{Korea Astronomy and Space Science Institute, 776 Daedeok-daero, Yuseong-gu, Daejeon 34055, Republic of Korea}

\author[0000-0002-6753-2066]{Mark R.~Morris}
\affiliation{Department of Physics and Astronomy, University of California, Los Angeles, CA 90095-1547, USA}

\author[0000-0002-1229-0426]{Jongsoo Kim}
\affiliation{Korea Astronomy and Space Science Institute, 776 Daedeok-daero, Yuseong-gu, Daejeon 34055, Republic of Korea}

\author[0000-0002-3938-4393]{Jinhua He}
\affiliation{Yunnan Observatories, Chinese Academy of Sciences, 396 Yangfangwang, Guandu District, Kunming, 650216, P.~R.~China}
\affiliation{Chinese Academy of Sciences South America Center for Astronomy, National Astronomical Observatories, CAS, Beijing 100101, P.~R.~China}
\affiliation{Departamento de Astronom\'{i}a, Universidad de Chile, Casilla 36-D, Santiago, Chile}

\correspondingauthor{Hyosun Kim}
\email{hkim@kasi.re.kr}

\begin{abstract}
We develop a physical framework for interpreting complex circumstellar patterns whorled around asymptotic giant branch (AGB) stars by investigating stable, coplanar triple systems using hydrodynamic and particle simulations. The introduction of a close tertiary body causes an additional periodic variation in the orbital velocity and trajectory of the AGB star. As a result, the circumstellar outflow builds a fine non-Archimedean spiral pattern superimposed upon the Archimedean spiral produced by the outer binary alone. This fine spiral can be approximated by off-centered circular rings that become tangent to each other at the location of the Archimedean spiral. The superimposed fine pattern fades out relatively quickly as a function of distance from the center of the system, in contrast to the dominant Archimedean spiral pattern, which presents a much slower fractional density decrease with radius. The different rates of radial decrease of the density contrast in the two superimposed patterns, coupled with their different time and spatial scales, lead to an apparent, but illusory radial change in the observed pattern interval, as has been reported, for example, in CW Leo. The function describing the detailed radial dependence of the expansion velocity is different in the two patterns, which may be used to distinguish them. The shape of the circumstellar whorled pattern is further explored as a function of the orbital eccentricity and the inner companion's mass. Although this study is confined to stable, coplanar triple systems, the results are likely applicable to moderately noncoplanar systems and open interesting avenues for studying noncoplanar systems.
\end{abstract}


\section{Introduction}\label{sec:int}

Planetary nebulae (PNe) and preplanetary nebulae (pPNe) exhibit a vast variety of morphologies, and spherical PNe are extremely rare. Only 3.4\% of 119 young PNe and none of 23 pPNe in the survey using the Hubble Space Telescope have a round shape, while the majority ($>60\%$) of them were classified as bipolar or multipolar \citep{zuc86,sah07,sah11}. The nonspherical morphologies, in general, include highly collimated jet-like features, bipolar compact knots, point-symmetric filaments (or strings of knots), bipolar ansae, and equatorially dense disk/torus-like features. Also invoked nowadays are objects displaying moderate deviations from spherical symmetry, such as spirals, intertwining off-centered circular rings, and irregularly distributed arcs, often found to be surrounding mass-losing asymptotic giant branch (AGB) stars and their compact descendants at the nuclei of pPNe and PNe. As noted in a review by \citet{bal02} and references therein, no single mechanism for shaping offers a comprehensive explanation of all observational properties of such systems, but it is nowadays commonly accepted that these structures require an interaction with a ``close'' binary companion \citep{dem09}.

A spiral-shell pattern surrounding a mass-losing giant star is an important tool for identifying binarity \citep[e.g.,][]{sok94,mas99,kim12b}. But the observed whorled patterns examined to date mostly, if not all, indicate ``wide'' (long-period) binaries. By searching $\sim650$ optical and infrared images of pPNe and PNe, \citet{ram16} found that 29 sources possess whorled patterns, and the time lapse between their consecutive rings and arcs ranges from 500--1200\,yr. The time intervals between successive waves in the circumstellar whorled patterns are also estimated using spatio-kinematic structures revealed in molecular line observations, including 350, 300, and 800\,yr for the well-known carbon-rich AGB stars (carbon stars), R Scl, CIT 6, and AFGL 3068, respectively \citep{mae12,kim13,kim17}. The (sub)millimeter images obtained in a new survey of oxygen-rich AGB stars seem to indicate the time interval to be not less than 30\,yr \citep{dec20}. Therefore, so far, the time interval of the circumstellar whorled pattern (corresponding to the orbital period of the binary) does not seem to indicate the presence of companions that are close enough to induce the morphological transitions from AGB to PN stages. This observational result is possibly due to a bias by observers who tend to target close, large sources by taking into account the limitations of telescopes in angular resolution and sensitivity. Another possibility, which we focus on in this paper, is that observers may have missed the spatial clues implying the presence of another (inner) companion in a triple system appearing as a minor pattern having less surface brightness (or density) contrast and a smaller pattern interval.

Classical statistical studies of stellar populations indicated that the fraction of triple to binary systems is about 0.11 and the fraction for higher multiplicity systems consisting of $n$ objects is $f_n=N_n/N_{n-1}\sim0.25$ \citep{duq91,tok01}. Thanks to high-resolution imaging, recent discoveries of additional subsystems in known binaries further boost the fraction of hierarchical multiplicity relative to binaries up to 30--50\% \citep{rag10,hir21}. This result implies that at least 3 out of 10 observed patterns whorled around evolved stars may present additional features induced by a tertiary component. In addition, close binaries tend to be accompanied by at least one other star, with the probability of finding additional companions increased by up to 96\% for spectroscopic binaries in the shortest-period group in \citet{tok06}. This naturally suggests that the progenitors of strongly bipolar PNe, which presumably originated from close binaries, may also possess an outer whorled pattern characterized by a longer time interval.

Before invoking a third stellar component, a binary model with a highly eccentric orbit can be considered to explain a bipolar outflow in terms of the short, but impactful, pericenter passages of the companion star.
As an example, the carbon star V Hya has experienced high-speed bullet-like ejections once every $\sim8.5$\,yr, which are hypothesized to be produced during the periapse passages of a binary companion having an orbital period of $\sim8.5$\,yr \citep{sah16}. However, \citet{sal19} raised an issue with the long-term orbital stability of such a binary system because the eccentric orbit should be circularized within a relatively short time by the dynamical and tidal interactions of the companion with the AGB star. They proposed that a triple system could be a solution for V Hya by invoking an orbital configuration in which the inner companion ($\la0.01\,\Msun$) grazes the Roche limit of the mass-losing star in an eccentric orbit, while the eccentricity of that orbit can be maintained by the gravitational influence of an outer companion.

The multiple-shell structure of the closest carbon star, CW Leo (a.k.a.\ IRC+10216), has been under debate since the discovery of the shells \citep{mau99}. The multiple shells appear roughly spherical, but they are not simply explained by equally spaced concentric circles; some of them are incomplete arcs, some are spiral-like, and some even appear to intertwine with other shells. Many authors also noticed that the centers of curvature are offset from each other \citep[e.g.,][see also \citealp{gue93}]{mau99,gue18}. By combining molecular line data taken with the Atacama Large Millimeter/submillimeter Array (ALMA; at a resolution of $\sim0\farcs3$) with those from the Submillimeter Array (SMA; at a resolution of $\sim3\arcsec$), \citet{gue18} found a regular interval of shells of $\sim16\arcsec$ (or 2000\,au at the distance of $\sim123$\,pc derived by \citealp{gro12}) in the outer envelope (up to a radius of $\sim110\arcsec$), which is consistent with a binary model having an orbital period of $\sim700$\,yr. Notice that \citeauthor{gue18} claimed that the orbit of the binary star system is eccentric and is viewed nearly face-on \citep[see also][]{cer15}. This geometry associated with a face-on and eccentric orbit binary was supported by a separate study of the position-angle dependence of the transverse wind velocity \citep{kim21,kim23}. However, this proposed geometry is contradictory to the nearly edge-on geometry proposed based on the elongated shape of the dust continuum emission and the bipolar-like optical image of the central 1\arcsec\ region \citep[e.g.,][]{men01,jef14,dec15}. \citeauthor{gue18}\ also noted that the spatial and time intervals between shells decrease in the inner (and younger) part of the envelope at radii $r<10\arcsec$, where $\Delta r\sim2\arcsec$ ($\Delta t<100$\,yr), compared to larger radii between 10\arcsec\ and 40\arcsec, where $\Delta r=5\arcsec$--10\arcsec\ ($\Delta t\sim300$\,yr). Their speculation that such a reduction of orbital period was caused by mass transfer either to the envelope or to the companion star fails because, as they argue, it requires a much higher mass-loss rate than the current rate of CW Leo. These discrepancies between the inner and outer parts of the circumstellar envelope may imply the presence of an additional (inner) companion.

In this paper, we present hydrodynamic and particle (kinematic) simulations for the circumstellar structure induced by a triple system consisting of a mass-losing star and two companion stars without mass loss. We mostly focus on particle simulations in order to explicitly track the circumstellar density-enhanced pattern and its velocity structure without the complexities due to gas pressure effects. Comparison of the particle simulation with the corresponding hydrodynamic simulation emphasizes the purely hydrodynamic effects, in particular, the density dispersion caused by shocks and gas pressure. Unlike hydrodynamic simulations, particle simulations can employ a constant velocity for the intrinsic wind of the mass-losing star, which facilitates the comparison of models over a wide parameter space for the stellar properties, e.g., the stellar masses. In a hydrodynamic simulation, even in a single star case, the wind velocities throughout the simulation domain are dependent upon the stellar masses (see Section\,\ref{sec:hdm}). Particle simulations also have an advantage in computational efficiency, with which we can trace the pattern out to the radii where the pattern induced by the third star has been repeated on multiple scales. This paper will also inform the merits and limits of particle simulations, which often have been, and will be, utilized for quick modeling of observed whorled patterns in the interpretational framework of binary (and hereafter, multibody) systems.

In Section\,\ref{sec:met} we describe our numerical methods for determining the orbits of a triple system (Section\,\ref{sec:orb}), for the hydrodynamic simulations (Section\,\ref{sec:hdm}), and for the pinwheel model based on following the trajectories of wind particles ejected at different moments (Section\,\ref{sec:pwm}). The results are shown in Section\,\ref{sec:res}. The implications of our models are discussed, and our conclusions are presented in Section\,\ref{sec:dis}.

\section{Numerical Methods}\label{sec:met}

\subsection{Coplanar Triple System}\label{sec:orb}

The preservation of dynamical stability within triple systems hinges upon the implementation of a hierarchical configuration, wherein an inner binary is orbited by an outer entity with a much wider orbital trajectory \citep{sal19}. Systems with smaller ratios of outer to inner orbital periods are susceptible to being unstable because of the eccentric Kozai--Lidov effect \citep{koz62,lid62}, unless the system is coplanar and the orbits are near-circular. Among evolved star systems that are suspected to have a third object, the AGB star $\pi^1$ Gru could be an example of a hierarchical triple with an orbital period ratio of $>500$, suggested by the recent finding of evidence for a close companion with an orbital period of only 10\,yr \citep{hom20}.
However, owing to the very large ratio between the orbital periods, the circumstellar patterns associated with the individual companions could have a large density contrast with respect to each other. It would likely leave a vestige of one of the companion stars on the circumstellar envelope that would be too tenuous to be observable, either because of the density attenuation of the larger pattern or because one companion is too low mass to build a significant pattern. As our aim in this work is to find and track the third object's footprints remaining in the circumstellar medium, as may have been observed in some AGB sources, like CW Leo, a large orbital period ratio is not addressed in our investigations.

Adopting such a framework, we assume a stable triple system, for which we further assume coplanar orbits. Coplanar orbits are relatively less affected by the eccentric Kozai--Lidov mechanism, albeit not completely free of such a mechanism, leading to large-amplitude eccentricity and inclination oscillations in near-coplanar triple systems \citep{li14}. The eccentric Kozai--Lidov mechanism is not taken into account in this paper because the timescales for its operation are much longer than the few-orbit timescales that we are considering here, and possibly even longer than the duration of the AGB phase.

Seven parameters that determine the orbits within a triple system include the masses of the three objects ($M_{\rm A}$, $M_{\rm B}$, and $M_{\rm C}$), the average separation between the center of mass of the inner (A--C) binary system and the outer B object ($a_{\rm AC}+a_{\rm B}$), the average separation between the inner objects ($a_{\rm A}+a_{\rm C}$), and their eccentricities ($e_{\rm AC-B}$ and $e_{\rm A-C}$). Here, $a$ represents the semimajor axis of the orbit of a star relative to the center of mass of the corresponding (AC--B or A--C) binary system.

We first calculate the orbits of a binary system composed of two mass components, $M_{\rm A}+M_{\rm C}$ and $M_{\rm B}$, separated by $a_{\rm AC}+a_{\rm B}$, which determines the final orbit of object B as well as the time-dependent position of the center of mass of the inner binary system, objects A and C. The mass ratio, $M_{\rm A}/M_{\rm C}$, and the A--C separation, $a_{\rm A}+a_{\rm C}$, then impose the individual orbits of objects A and C with respect to the time-dependent position of the center of mass of the inner binary system.

\begin{figure*} 
  \plotone{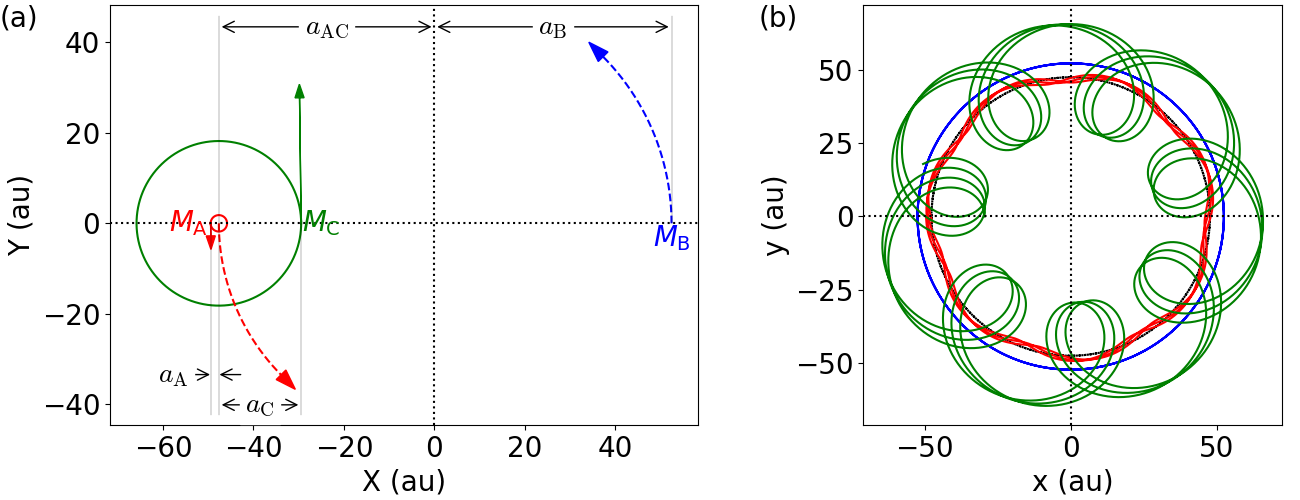}
  \caption{\label{fig:orb}
    Orbits of three objects in a triple system used for the calculations
    displayed in Figures\,\ref{fig:den} and \ref{fig:vel}. All three stars
    are initially aligned along the $x$-axis in the order of A--C--B from
    left to right, starting from their apocenters in eccentric orbit cases.
    (a) Orbits of objects A, B, and C with individual masses of $M_{\rm A}$,
    $M_{\rm B}$, and $M_{\rm C}$ in the corotating $XY$ frame in which the
    locations of object B and the center of mass of the A--C binary system
    are fixed. The center of mass of the triple system is located at the
    origin. The term $a_{\rm AC}$ denotes the (average) distance of the
    center of mass of the A--C binary system from the center of mass of
    the whole system, while $a_{\rm B}$ indicates that of object B.
    Dashed curves schematically indicate the directions of the motions
    of the center of mass of the A--C binary system (red) and of object
    B (blue) in the observer's frame.
    The notations $a_{\rm A}$ and $a_{\rm C}$ demonstrate the (average)
    distances of objects A and C, respectively, with respect to the center
    of mass of the A--C binary system. Red and green circles present the
    orbits of A and C in this rest frame. Red and green solid straight
    lines display the velocity vectors of objects A and C at their initial
    positions, and are scaled to each other. (b) Orbits of objects A (red),
    B (blue), and C (green) in the frame of nonrotating observers located
    on the $+z$-axis.
  }
\end{figure*}

Figure\,\ref{fig:orb} illustrates the orbits of a triple system (a) in $XY$ coordinates corotating with object B about the center of mass of the three objects (coinciding with the coordinate origin) and (b) in the inertial frame of observers sitting on the $+z$-axis. In the former frame, the locations of the center of mass of the inner binary system and object B are fixed, and the orbits of A and C are closed and circular (elliptical, in eccentric orbit cases). In the observer's frame, however, the orbits of A and C are not necessarily closed. The small fluctuation in the orbit of the mass-losing star A (red curve in Figure\,\ref{fig:orb}(b)), relative to a circle, is the primary cause of the main features described in this paper.

This study is limited to triple systems having coplanar orbits in order to capture the essential features created by the presence of a third object. Finally, the orbital motions are assumed to be stable in their predefined circular or eccentric orbits, with constant orbital parameters. In future studies, stability considerations will be included in the modeling of the observations.

\subsection{Hydrodynamic Simulations}\label{sec:hdm}

Three-dimensional hydrodynamic simulations for triple systems were performed using version 4.5 of the code FLASH \citep{fry00} by solving the Eulerian hydrodynamic equations in Cartesian coordinates with the origin at the center of mass of the triple system. The adaptive mesh refinement scheme imposes a higher level of refinement toward the instantaneous stellar positions. The refinement is determined by the stellar positions, not by the usual density or velocity criteria.
Specifically, the maximum refinement level of 9, with 64 grid points per block length for a total image size of 6000\,au, imposes the highest resolution of about 0.37\,au near the stellar positions; it assigns 19 grid points to the diameter (7\,au) of the hypothesized wind-launching sphere surrounding the mass-losing star, where the initial vector velocities in the observer's frame of the wind material blowing out of the star are reset every simulation time step. The mass-losing star is treated as a point source having a gravitational softening radius (also called the Plummer radius) of 1\,au. The mass-loss rate for all models presented in this paper has been set at $3\times10^{-6}\,\Mspy$; the morphological results are insensitive to the assumed mass-loss rate. The equation of state of an ideal gas is assumed with the ratio of specific heats, $\gamma$, chosen to be equal to 1.4. Radiative cooling and heating are not included for simplicity; we only follow the density of the material.

Following a commonly adopted technique for wind acceleration \citep[e.g.][]{the93}, the gravitational force attributed to the mass-losing star of mass $M_{\rm A}$ located at $\vct{r} = \vcttwo{r}{A}$ is reduced by a factor of $1-f$, where the wind acceleration factor $f$ is a constant representing the ratio of the outward force due to radiation pressure on dust grains to the inward gravitational force (see the wind velocity profiles for different constant values of $f$ in \citealp{kim12a} for an isothermal gas; the wind solution for a polytropic gas can be found in \citealp{shi21}). In a test simulation for a corresponding single mass-losing star located at the center of the simulation domain, the intrinsic wind profile as a function of radius followed a supersonic branch of the above papers as the updated version of Parker's wind solution \citep{par58}, and the terminal velocity was around 13\,\kmps. The intrinsic wind velocity is governed by the momentum equation of hydrodynamics; therefore, it would be scaled down with an increased companion mass.

The gravitational forces attributed to the outer and inner companion stars of mass $M_{\rm B}$ and $M_{\rm C}$ located at $\vct{r} = \vcttwo{r}{B}$ and $\vct{r} = \vcttwo{r}{C}$, respectively, are also implemented in the code. In order to explore the effects of the perturbed orbital path of the mass-losing star and to distinguish those effects from those created by the density wakes formed behind the companion stars, we run simulations that alternatively include and exclude the terms for the companions' gravitational forces by setting the shutoff parameter, $\mathcal{W}$, defined by \citet{kim19} and \citet{kim23}, to 1 and 0, respectively. In any case, the orbit of the mass-losing star is preset according to the dynamics of the triple system, as described in Section\,\ref{sec:orb}. The results of the hydrodynamic simulations are displayed in Figures\,\ref{fig:den}(a) and (b) for density and in Figures\,\ref{fig:vel}(a) and (b) for the expansion velocity; a further description can be found in Section\,\ref{sec:res}.

\subsection{Particle Simulations}\label{sec:pwm}

We consider that most of the characteristics of the spiral-shell pattern induced by a binary or triple system can be explained by a simpler model that tracks the trajectories of wind particles ejected from the mass-losing giant star. In order to identify the purely hydrodynamic effects, we compare the density and velocity fields obtained through a hydrodynamic simulation with the corresponding particle model characteristics calculated by the pinwheel code, as described below.

The first version of the pinwheel model accumulates the particles that are freely moving with the instantaneous momenta gained at the moments of their ejection from the mass-losing star as it passes through the predefined positions along its orbit with the orbital velocities as calculated in Section\,\ref{sec:orb}. Here, the particle velocity is defined as the vector sum of the intrinsically isotropic wind velocity (denoted by \Vexp) with the orbital velocity of the star.
If $N_{\rm par}$ particles are ejected with equal spacing over all solid angles, from one of the discrete $N_{\rm pos}$ positions along the orbit of the star within one orbital period, then within the observing time $t=N_{\rm turn}$ orbits, a total of $N_{\rm par} \times N_{\rm pos} \times N_{\rm turn}$ particles are moving within the computational domain. Among them, the number of particles located within each grid cell is used as a measure of the density distribution. The velocity map is also calculated by averaging the velocity vectors of individual particles falling into each grid cell. An example binary model that was calculated by this method is shown in \citet{kim17}'s supplementary Figure\,4.

The second version of the pinwheel code is written using the concept of a piston based at the systemic center of mass. Once the direction of a piston $(\theta,\,\phi)$ is chosen, with the usual notations $\theta$ for the polar angle and $\phi$ for the azimuthal angle, the time-variable length and velocity of the piston head are determined by the projection of the position and velocity of the star in its orbit onto the piston direction:
\begin{eqnarray}
  l_{\rm piston} &=& (x_{\rm A}\cos\phi + y_{\rm A}\sin\phi) \sin\theta,\\
  V_{\rm piston} &=& (v_{x,\rm A}\cos\phi + v_{y,\rm A}\sin\phi) \sin\theta,
\end{eqnarray}
where $(x_{\rm A},\,y_{\rm A})$ and $(v_{x,\rm A},\,v_{y,\rm A})$ indicate the position and velocity of object A in the orbital plane. The wind particle is ejected with speed $\Vexp+V_{\rm piston}$ toward the piston direction. The output images of the first and second versions of the pinwheel model are apparently the same under the condition of a far-distance approximation (i.e., $r\gg l_{\rm piston}$).

\begin{deluxetable*}{cccccccccccc}
  \tablecaption{\label{tab:par}%
    Parameters for particle simulations}

  \tablehead{
    \colhead{$M_{\rm A}$} & \colhead{$M_{\rm B}$} & \colhead{$M_{\rm C}$}
    & \colhead{$a_{\rm AC}+a_{\rm B}$} & \colhead{$a_{\rm A}+a_{\rm C}$}
    & \colhead{$e_{\rm AC-B}$} & \colhead{$e_{\rm A-C}$} & \colhead{Stickiness}
    & \colhead{$N_{\rm pos}$}&\colhead{$N_{\rm par}$}&\colhead{$N_{\rm turn}$}
    & \colhead{Density}\\[-1ex]
    \colhead{($\Msun$)} & \colhead{($\Msun$)} & \colhead{($\Msun$)}
    & \colhead{(au)} & \colhead{(au)}
    & \colhead{} & \colhead{} & \colhead{(\% or \kmps)}
    & \colhead{} & \colhead{} & \colhead{} & \colhead{(Figures)}}

  \startdata
  \tableline
  1&1&0.1 & 100&20 & 0&0 & 0\% & 2000 & 2000 & 3 & \ref{fig:den}(c)\\
  1&1&0.1 & 100&20 & 0&0 & 100\% & 2000 & 2000 & 3 & \ref{fig:den}(d)\\
  \tableline
  1&1&0.1 & 100&20 & 0&0 & 0.5 & 1000 & 2000 & 5 & \ref{fig:ecc}(a)\\
  1&1&0.1 & 100&20 & 0&0.8 & 0.5 & 1000 & 2000 & 5 & \ref{fig:ecc}(b)\\
  1&1&0.1 & 100&20 & 0.8&0 & 0.5 & 1000 & 2000 & 5 & \ref{fig:ecc}(c)\\
  1&1&0.1 & 100&20 & 0.8&0.8 & 0.5 & 1000 & 2000 & 5 & \ref{fig:ecc}(d)\\
  \tableline
  0.8&1&0.3 & 100&20 & 0&0 & 0.5 & 1000 & 2000 & 5 & \ref{fig:m_c}(a)\\
  1.09&1&0.01 & 100&20 & 0&0 & 0.5 & 1000 & 2000 & 5 & \ref{fig:m_c}(b)\\
  \enddata

  \tablecomments{In all models, the orbital periods are set the same as
    $T_{\rm AC-B}=690$\,yr and $T_{\rm A-C}=85$\,yr, yielding a ratio
    of $\sim8$. The gravitational wakes of the companion objects, B and
    C, are not considered in particle simulations, corresponding to the
    shutoff parameter $\mathcal{W}=0$ in the hydrodynamic simulations.}
\end{deluxetable*}

In the third version of the pinwheel code, the particles are made to be sticky. For this, a piston approximation is used for convenience. Whenever the particles ejected at later times overtake previously ejected particles, the velocities of all these particles at the same distance are averaged and updated to a common new value. This assumes that the particles spatially coinciding with each other perfectly stick together. A binary model of this version was exhibited in \citet{he07}.

The fourth version of the code reduces the efficiency of stickiness so that the velocity dispersion of the particles that are coincident with each other is reduced to the predefined speed of sound; if the initial velocity dispersion of the overlapping particles is smaller than the specified speed of sound, their velocities are not updated.

Figures\,\ref{fig:den}(c) and \ref{fig:vel}(c) present the resulting density and velocity maps of the first (or second) version of the pinwheel model (i.e., without stickiness), while Figures\,\ref{fig:den}(d) and \ref{fig:vel}(d) show the third version's results for particles with the efficiency of stickiness set to 100\%. The example density maps produced via the fourth version of the piston model with the velocity dispersion of 0.5\,\kmps\ for the coincident particles are exhibited in Sections\,\ref{sec:ecs} and \ref{sec:ecc}. Table\,\ref{tab:par} summarizes the parameters for the particle simulations displayed in this paper.

\section{Results}\label{sec:res}

\subsection{Hydrodynamic Simulation: Understanding the Roles of Individual Stars}\label{sec:hdr}

\begin{figure*} 
  \plotone{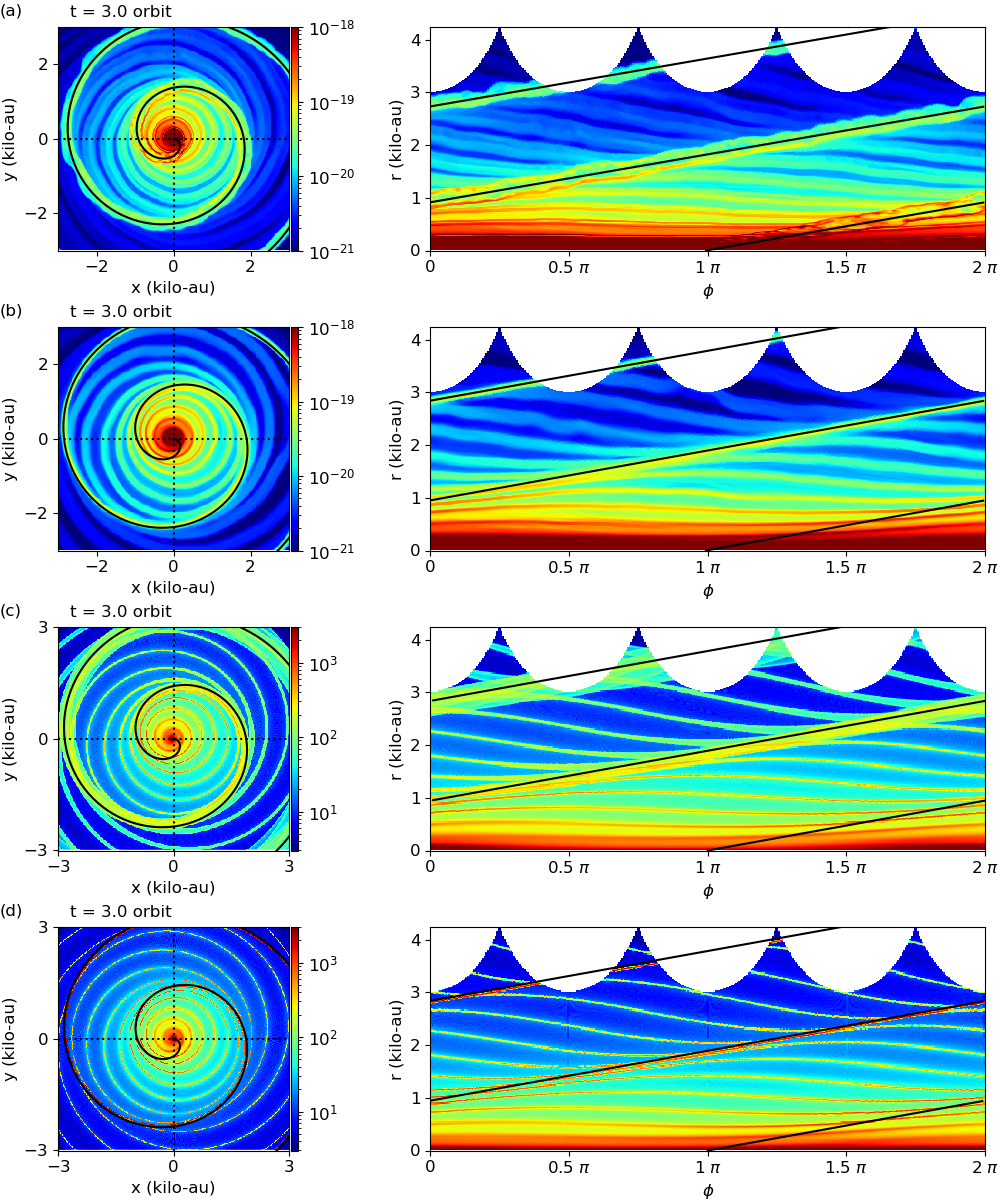}
  \caption{\label{fig:den}
    (a) Hydrodynamic simulation for the circumstellar matter distribution
    governed by orbital motions of three objects (with orbital parameters of
    $M_{\rm A}=1.0\,\Msun$, $M_{\rm B}=1.0\,\Msun$, $M_{\rm C}=0.1\,\Msun$,
    $a_{\rm AC}+a_{\rm B}=100$\,au, $a_{\rm A}+a_{\rm C}=20$\,au,
    $e_{\rm AC-B}=0$, and $e_{\rm A-C}=0$). The density in the orbital plane
    is displayed in a logarithmic scale in units of \mbox{g\,cm$^{-3}$}.
    The black solid line represents an Archimedean spiral with a pattern
    speed of 12.5\,\kmps. The angle $\phi$ is measured from the $-x$-axis
    in the clockwise direction.
    (b) Same as (a), but with the gravitational density wakes of objects B and
    C excluded by turning off their gravitational effects on circumstellar gas
    by setting the shutoff parameter, $\mathcal{W}$, to 0. The black solid line
    corresponds to an Archimedean spiral with a pattern speed of 13\,\kmps.
    (c) A pinwheel model with a constant wind velocity of $\Vexp=13\,\kmps$,
    adopted to match the major spiral pattern with that in panel (b), and (d)
    the corresponding piston model with sticky particles. The color bar in
    (c) and (d) panels shows the number counts of particles within each grid
    cell, which is a function of the total number of particles present in the
    computational domain after the computation has reached a steady state.
  }
\end{figure*}

Figure\,\ref{fig:den}(a) shows the density distribution in the (common) orbital plane of the triple star system with the mass-losing giant star of mass $M_{\rm A}=1\,\Msun$, the outer companion of mass $M_{\rm B}=1\,\Msun$, and the inner companion of mass $M_{\rm C}=0.1\,\Msun$. The distances from the giant star are 100\,au and 20\,au for the outer and inner companions, respectively, and the orbital shapes are perfectly circular.

We note that a binary system consisting of two objects with masses $M_{\rm A}+M_{\rm C}$ and $M_{\rm B}$ induces the density-enhanced structure of an Archimedean spiral in the orbital plane, as denoted by the black line in the figure, satisfying
\begin{equation}\label{eqn:slp}
  r/\phi = \Vexp\, T_{\rm AC-B} / 2\pi,
\end{equation}
where $T_{\rm AC-B}$ represents the orbital period in this binary system.
Figure\,\ref{fig:den}(a) shows that the triple system creates the same spiral structure as in the abovementioned binary system (hereafter referred to as the main spiral), along with finer structures in the inter-ridge region of the main spiral.
As shown in the right panel for the $r$--$\phi$ polar coordinate map, the finer structures have a spiral shape with the opposite orientation (i.e., the slope in the polar coordinate map is negative). The segments of the finer spiral structure effectively merge into the main spiral ridge, rather than passing continuously across it.

The overall features in Figure\,\ref{fig:den}(a) are compared to the correspondences in Figure\,\ref{fig:den}(b), where the gravitational wakes of objects B and C are excluded; we have further compared it to the images (not shown in this paper) made by excluding just one of the gravitational wakes of objects B and C.
The nonlinear perturbations appearing along the main spiral, as clearly seen in both the left and right panels in Figure\,\ref{fig:den}(a), can be attributed to the overlay of the wake of the outer companion (object B). The irregularities in the main spiral have a vertically limited extent, as indicated in previous studies \citep[see, e.g.,][]{kim12a,kim12b,kim19}.
On the other hand, the broadening of the fine spiral segments shown in Figure\,\ref{fig:den}(a), relative to the width of the corresponding features shown in Figure\,\ref{fig:den}(b), occurs with the introduction of the gravitational effect of the inner companion (object C). The density profile across the fine spiral is also modified from two peaks shown in Figure\,\ref{fig:den}(a) to one peak in Figure\,\ref{fig:den}(b).
We also note that the slope of the black line in the right panel is greater in Figure\,\ref{fig:den}(b) than in Figure\,\ref{fig:den}(a) because of the absence of the companions' gravitational influences in Figure\,\ref{fig:den}(b), which yields an expansion velocity that is about 0.5\,\kmps\ greater than in Figure\,\ref{fig:den}(a). The deceleration of the wind is smaller when the companions' gravity is excluded.

With increasing distance from the system's center of gravity, the relative density decrease of the finer pattern is much faster than that of the main spiral. Furthermore, while the radial broadening of the main spiral is insignificant, the width of the finer pattern increases with radius, eventually smoothly filling the region between the density ridges of the main spiral. Therefore, the finer pattern in the outermost part would likely be unidentified in, for example, molecular line observations of the circumstellar medium of an AGB star in a triple system.

In Figure\,\ref{fig:vel} for the distribution of expansion velocity, the velocity gradient is maximized at the location of the main spiral pattern, traced by the black solid line. Overall, the expansion velocity gradually increases from one arm to the next outer arm of the main spiral. The velocity jumps at the locations of fine spiral structures are minor, which provides an important characteristic to distinguish the main and fine spirals of a triple system among the observed intensity peaks.

\subsection{Particle Simulations: Differentiating Hydrodynamic and Nonhydrodynamic Effects}\label{sec:pwr}

Panels (c) and (d) of Figure\,\ref{fig:den} present the results of pinwheel model calculations with stickiness efficiencies of 0\% and 100\%, respectively. Both images well approximate the locations of the whorled patterns observed in the hydrodynamic model displayed in Figure\,\ref{fig:den}(b). The width of the main spiral pattern in the hydrodynamic simulation is, however, smaller than in the nonsticky model shown in Figure\,\ref{fig:den}(c) and larger than in the extremely sticky model in Figure\,\ref{fig:den}(d). It suggests the necessity of adjusting for the efficiency of the stickiness of the particles when they coincide.

\begin{figure*} 
  \plotone{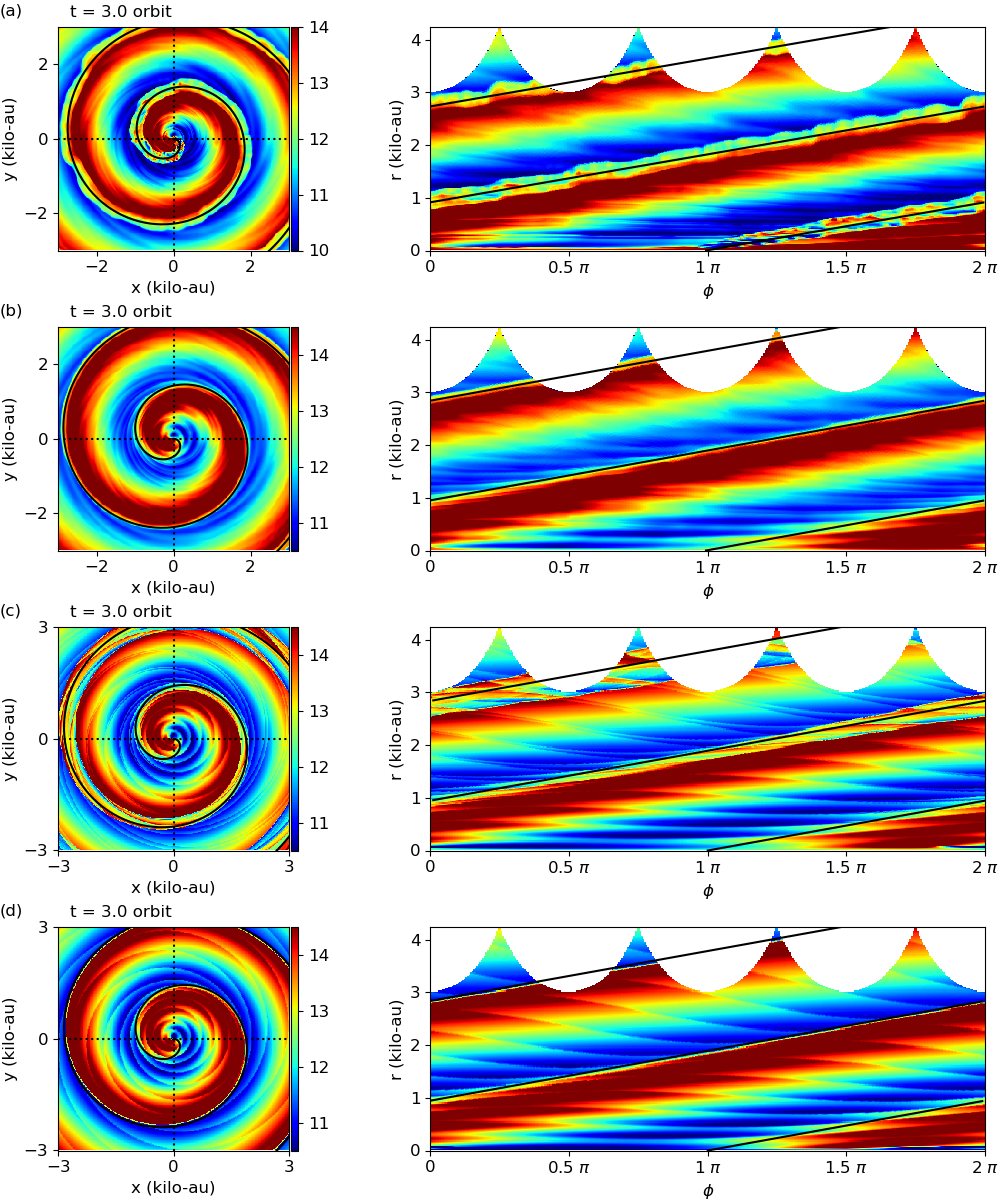}
  \caption{\label{fig:vel}
    Same as Figure\,\ref{fig:den}, but illustrating the expansion velocity
    of the fluid in the orbital plane.
  }
\end{figure*}

The net velocity distribution, as plotted in Figure\,\ref{fig:vel}, is also overall well reproduced by the pinwheel model, in particular with the sticky particles. In the perfectly sticky model shown in Figure\,\ref{fig:vel}(d), the expansion velocity sharply drops from $>14$ to $<11\,\kmps$ at the main spiral's ridge, whose width is unresolved in this model. The corresponding hydrodynamic model, drawn in Figure\,\ref{fig:vel}(b), has a similar velocity distribution, except for the slightly larger width of the ridge of the main spiral. In the nonsticky model, the relatively smaller velocity between the split edges of the main spiral is distributed over a wider area (see Figure\,\ref{fig:vel}(c)), inconsistent with the hydrodynamic result shown in Figure\,\ref{fig:vel}(b). Within the region between the split edges of the main spiral, it is also found that the continuation of the sharp velocity structures of the finer pattern having the opposite orientation does not cross the black solid line (see the right panel of Figure\,\ref{fig:vel}(c)).

Although there is a close similarity in the global distribution of the expansion velocity of fluid, the number of fine spirals in the inter-ridge region of the main spiral is doubled in the hydrodynamic model (Figure\,\ref{fig:vel}(b)) compared to that in the pinwheel model with sticky particles (Figure\,\ref{fig:vel}(d)). The fine spirals have an extremely small width in the sticky model, while they show broadening in both the nonsticky and hydrodynamic models, accompanied by double-peaked velocity profiles (\citealp[see Figure\,6 of][]{kim19}, and below for more details).

In the full three-body hydrodynamic model, the nonlinearities in the density wake of the outer companion are apparent in Figure\,\ref{fig:vel}(a). The fine spirals in the full three-body model are broad (compare Figures\,\ref{fig:den}(a) and (b)), within which triple peaks are presented, owing to the overlay of the gravitational wake of the inner companion. Accordingly, the velocity structures of the fine partial spirals are more complex (compare Figures\,\ref{fig:vel}(a) and \ref{fig:vel}(b)).

\begin{figure*} 
  \plotone{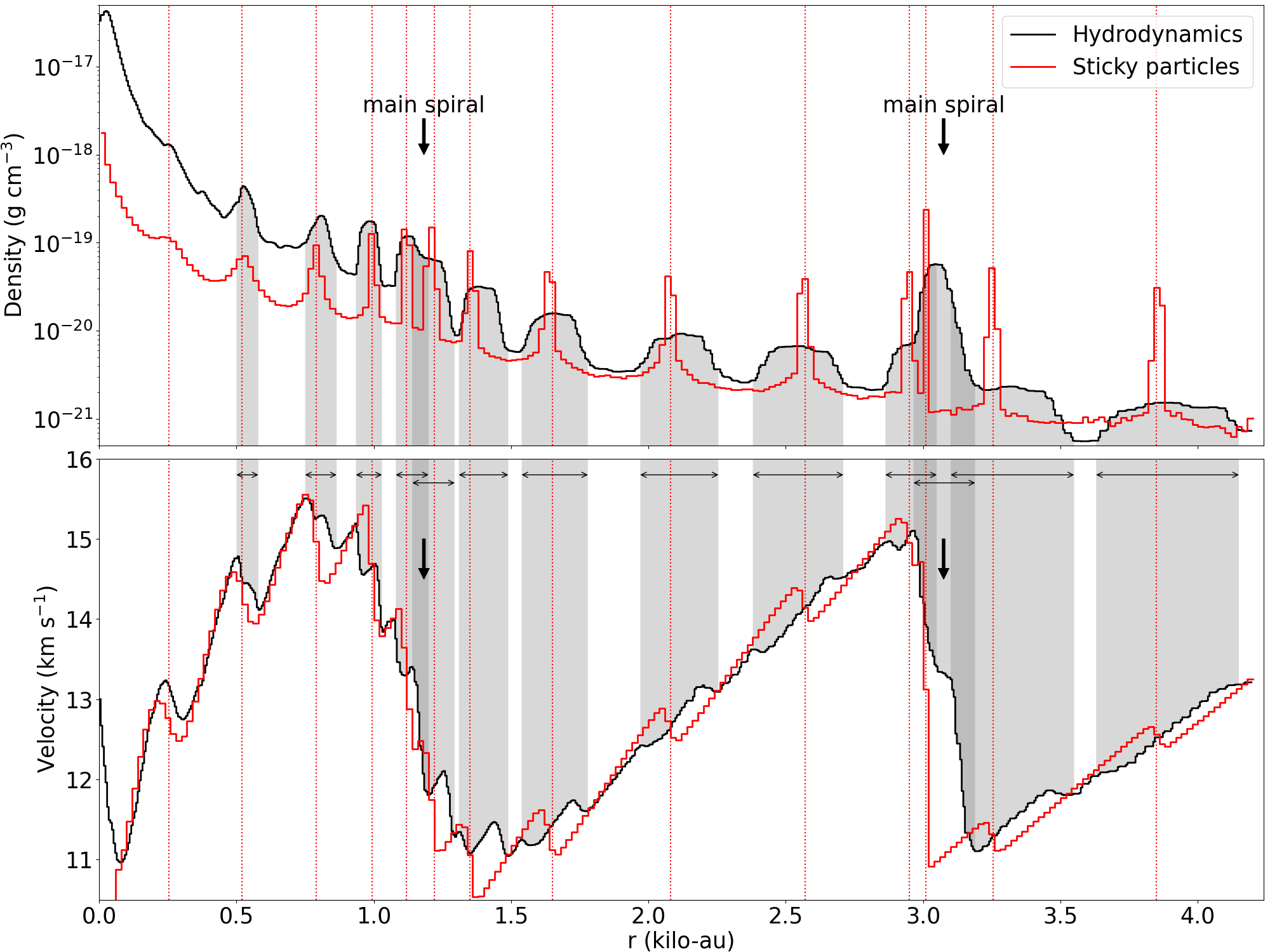}
  \caption{\label{fig:prf}
    Comparison between hydrodynamic and sticky particle models in density
    and velocity profiles. The density profiles along the azimuthal angle
    $\phi=\pi/4$, passing $(x,\,y)=(-1,\,1)$, in Figures\,\ref{fig:den}(b)
    and (d) are drawn in black and red curves, respectively, in the top
    panel. The corresponding velocity profiles from Figures\,\ref{fig:vel}(b)
    and (d) are plotted in the bottom panel. Red-dotted vertical lines mark
    the peak positions of the sticky particle model. Shaded regions, along
    with the horizontal two-headed arrows, identify the characteristic spiral
    ridges \citep[see Figure 6 of][and the relevant text therein]{kim19}.
  }
\end{figure*}

Figure\,\ref{fig:prf} compares the density and velocity profiles between the 100\% sticky particle model and the hydrodynamic model induced by the orbital motion of the mass-losing object. The red-dotted vertical lines indicate the positions of the peaks in the sticky particle model. Note that the typical radial structure of one spiral ridge is characterized by a one-peak density profile, surrounded by double-peaked temperature profiles, and the velocity profile with one inflection point \citep[see Figure\,6 of][]{kim19}. Each of the radial zones indicated by a horizontal two-headed arrow and shading in Figure\,\ref{fig:prf} shows that the dispersed spiral ridge of the hydrodynamic model coincides with a common shape of the velocity profile. Furthermore, the density peaks in the sticky particle model coincide with inflection points in the velocity profile, as indicated by the red-dotted vertical lines. The darker regions represent the overlaps of these shaded regions. These shaded regions, determined based on the velocity profile, agree well with the enhanced regions in the density profile.

The spirals of the hydrodynamic model are enclosed by two shocks at their inner and outer edges.  Their radial variations in density and velocity are very similar to the characteristic density and velocity profiles of an outgoing forward shock and a reverse shock, the latter of which accelerates material inward, relative to the expanding forward shock, as in a supernovae remnant \citep[e.g., Figure 1 in][]{tru99}, albeit showing smaller shock jumps due to a much slower wind speed. The individual peak structures of the fine spiral, shaded in Figure\,\ref{fig:prf}, have such a velocity profile, including one inflection point, superimposed upon the larger-scale velocity variation following the main spiral pattern. As a consequence, the density peaks in the hydrodynamic model are largely dispersed through the regions enclosed by the forward and reverse shocks at the inner and outer edges of the spiral patterns.

\subsection{Efficiency of the Stickiness of Wind Particles}\label{sec:ecs}

The efficiency of stickiness is adjusted in the piston model by allowing a certain range of velocity dispersion of the coincident particles up to a predefined constant value. As a result, the width of the main spiral in the hydrodynamic model is reasonably reproduced by applying a velocity dispersion of $\sim0.5\,\kmps$ to the pinwheel model (see Figure\,\ref{fig:ecc}(a)), which corresponds to the adiabatic speed of sound of a gaseous medium at a temperature of $\sim20$\,K. In our hydrodynamic simulations, the temperature shows an overall decrease with radius except for the jumps at the major and minor spiral patterns, and the interarm temperature is below 20\,K beyond 1 kilo-au.

\subsection{Eccentric Orbits}\label{sec:ecc}

\begin{figure*} 
  \plotone{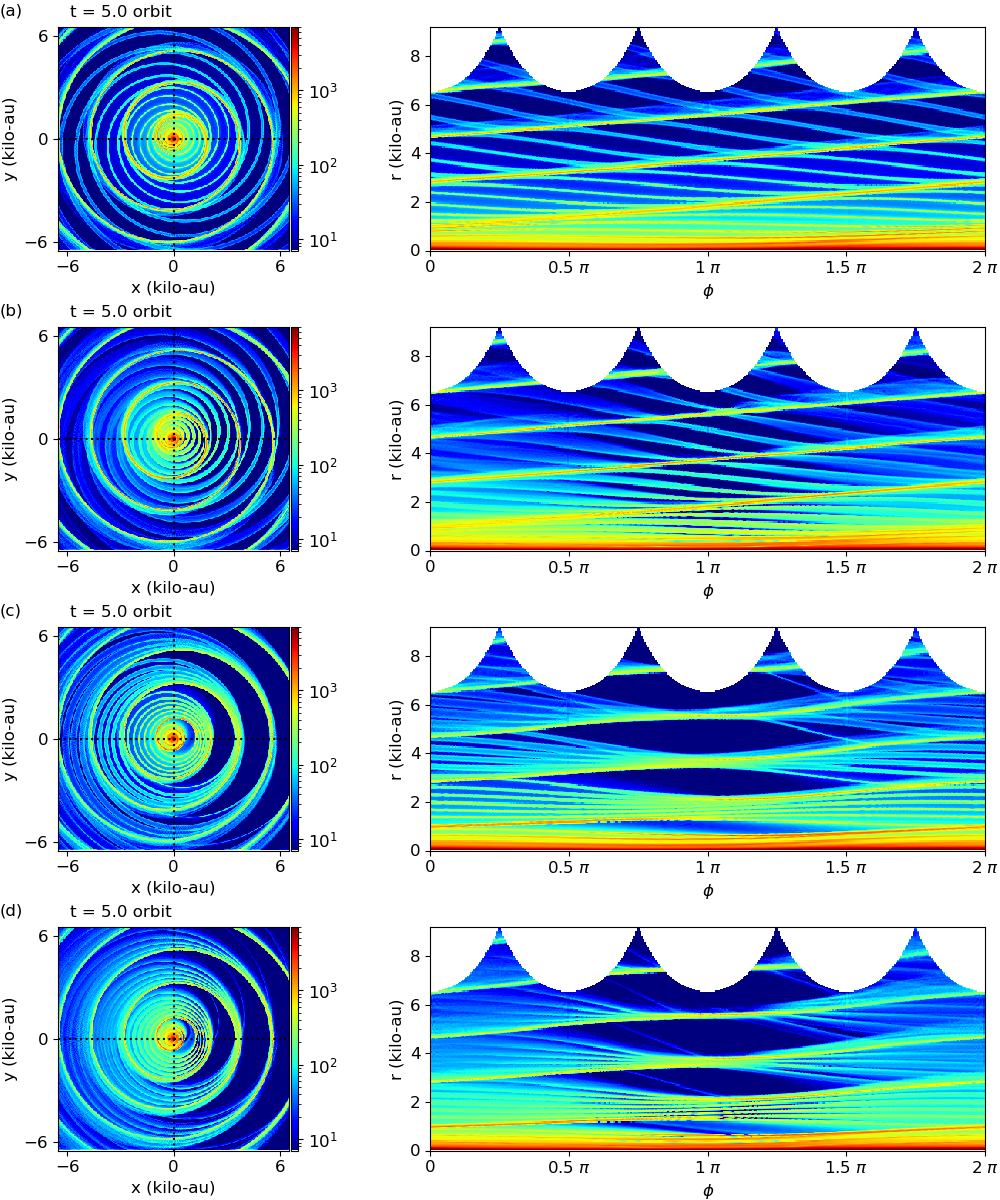}
  \caption{\label{fig:ecc}
    Midplane density in pinwheel models of a three-body system
    in which the mass-losing star is object A. The stickiness of
    particles is adjusted to provide a speed of sound of 0.5\,\kmps.
    The employed parameters are $\Vexp=13\,\kmps$,
    $M_{\rm A}=1.0\,\Msun$, $M_{\rm B}=1.0\,\Msun$, $M_{\rm C}=0.1\,\Msun$,
    $a_{\rm AC}+a_{\rm B}=100$\,au, and $a_{\rm A}+a_{\rm C}=20$\,au.
    The orbital eccentricity of object B ($e_{\rm AC-B}$) is 0 in the upper
    two models ((a) and (b)) and 0.8 in the bottom two models ((c) and (d)).
    The orbital eccentricity of objects A and C ($e_{\rm A-C}$) with respect
    to their center of mass is 0 in (a) and (c) but 0.8 in (b) and (d).
  }
\end{figure*}

In this section, we demonstrate the influences of eccentricities of stellar orbits in the density distribution of the circumstellar spiral patterns. The velocity dispersion up to $\sim0.5\,\kmps$ for the coincident particles is adopted.
Figure\,\ref{fig:ecc}(a) presents the case in which the orbital shapes of the three stars are all circular, as described in the previous sections. 
Compared to this, in the model with an eccentric orbit for the inner companion ($e_{\rm A-C}>0$, Figure\,\ref{fig:ecc}(b))\footnote{In reality, the extremely eccentric ($e=0.8$) inner companion that we adopted would be susceptible to Roche-lobe overflow, as the distance to the $L_1$ Lagrangian point, $\sim3$\,au, could be similar to, or smaller than, the radius of an AGB star. However, our simulations do not take this effect into account, which treated the mass-losing star as being volumeless.}, the individual patterns are similar in location but become very widened, in particular toward the apocenter of the mass-losing star ($\phi\sim0$, which we have defined to be in the $-x$ direction).

In the model with an eccentric orbit for the outer companion ($e_{\rm AC-B}>0$, Figure\,\ref{fig:ecc}(c))\footnote{According to Equation (1) of \citet{tok21}, a highly eccentric orbit of the outer companion would be unstable. We choose an unrealistically large eccentricity for the outer orbit for the purpose of maximizing the visual appearance of the pattern changes caused by orbital eccentricity.}, a one-sided dearth of matter in the inter-ridge regions is clearly seen toward the pericenter of the mass-losing star ($\phi=\pi$, or in the $+x$ direction), just like in a typical eccentric binary system \citep{kim15,kim19}. The fine spirals accordingly congregate near the main spiral at the position angle corresponding to the pericenter of the mass-losing star, making a bowtie shape in the $r$--$\phi$ map. The knot of the bowtie occurs at values of $\phi$ corresponding to the pericenter of the mass-losing star, and the knot is tighter for larger orbital eccentricities of the outer companion. In this model, the broadening of the fine spiral pattern is nearly independent of the position angle.

Figure\,\ref{fig:ecc}(d) shows eccentric orbits for both inner and outer companions, in which the fine spirals are threaded within the bowtie (as in Figure\,\ref{fig:ecc}(c)), and the individual fine spirals are widened around $\phi\sim0$ (as in Figure\,\ref{fig:ecc}(b)).

\subsection{Mass of the Inner Companion}\label{sec:inn}

\begin{figure*} 
  \plotone{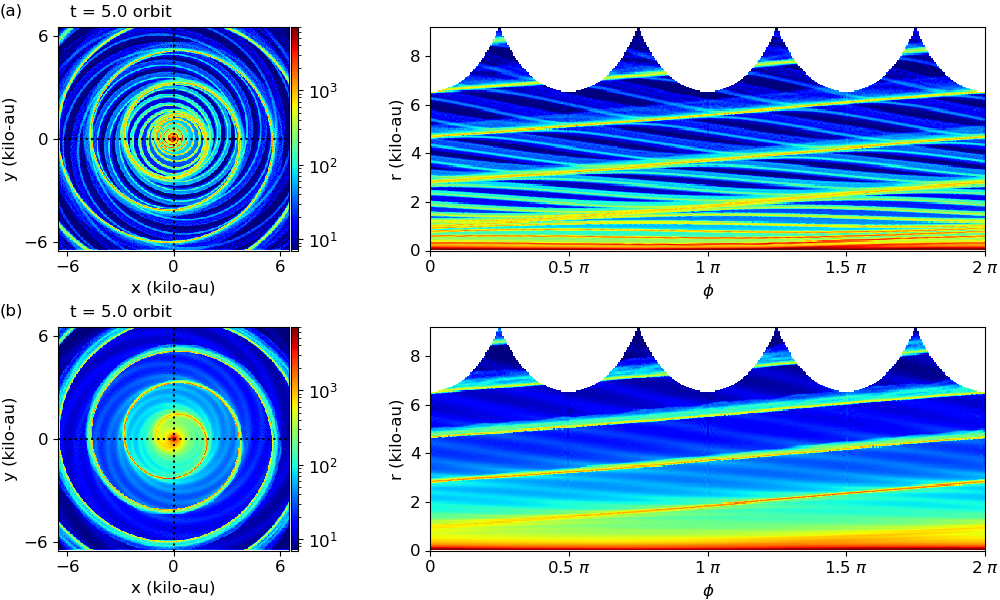}
  \caption{\label{fig:m_c}
    Same as in Figure\,\ref{fig:ecc}(a), but for
    (a) $M_{\rm A}=0.8\,\Msun$ and $M_{\rm C}=0.3\,\Msun$, while in
    (b) $M_{\rm A}=1.09\,\Msun$ and $M_{\rm C}=0.01\,\Msun$.
    Notice that $M_{\rm A}+M_{\rm C}$ is equivalent to the value for
    the models in Figure\,\ref{fig:ecc}; therefore, the orbital motions
    of object B and of the center of mass of objects A and C remain
    the same as in Figure\,\ref{fig:ecc}(a).
  }
\end{figure*}

The effect of the mass ratio $M_{\rm A}/M_{\rm C}$ at a given $M_{\rm A}+M_{\rm C}$ is also explored. We compared our fiducial model for the inner companion as a low-mass star ($M_{\rm C}=0.1\,\Msun$, Figure\,\ref{fig:ecc}(a)) with the model for a slightly more massive star companion ($M_{\rm C}=0.3\,\Msun$, Figure\,\ref{fig:m_c}(a)) and the model for a planetary-mass companion ($M_{\rm C}=0.01\,\Msun$, Figure\,\ref{fig:m_c}(b)). In the pinwheel model at a constant wind velocity of $13\,\kmps$, the fixed $M_{\rm A}+M_{\rm C}$ guarantees the same slope of the main spiral pattern. The mass $M_{\rm A}$ is 1.0, 0.8, and 1.09\,\Msun, respectively, in Figure\,\ref{fig:ecc}(a), Figure\,\ref{fig:m_c}(a), and Figure\,\ref{fig:m_c}(b).

From the comparison of the three cases for the mass of the inner companion, we find that the peak density of the pattern does not differ much, but the density in the inter-ridge regions is significantly reduced as the mass of the inner companion is increased, thereby increasing the density contrast of the pattern. The fine pattern formed due to the presence of a more massive inner companion is somewhat broader in width, which yields a countereffect in the density contrast because the column density in the finer spiral features becomes more spread out. The latter effect, governed by the velocity dispersion of the pattern, is, however, largely limited by the local speed of sound, which would decrease with increasing distance from the star. Because the speed of sound in the outer circumstellar envelope is likely to be lower than the single value used in the current simple calculations, the latter (widening) effect would be reduced in the outer circumstellar envelope, thereby preserving the high-density contrast. Therefore, the additional pattern established by a relatively massive inner companion tends to have a higher-density contrast, on top of the predominant binary-induced pattern.

\subsection{Circular Ring Approximation}\label{sec:cir}

The fine spirals are formed from the introduction of the inner companion (object C) into the binary system composed of the mass-losing star (object A) and the outer companion (object B). The actual cause of these spirals is the wiggling of the orbital trajectory of the mass-losing star (the red curve in Figure\,\ref{fig:orb}(b)) and the consequent variation of its orbital velocity. The wiggling of the orbit around the center of mass of the objects A and C is related to their orbital period, $T_{\rm A-C}$. 

\begin{figure*} 
  \plotone{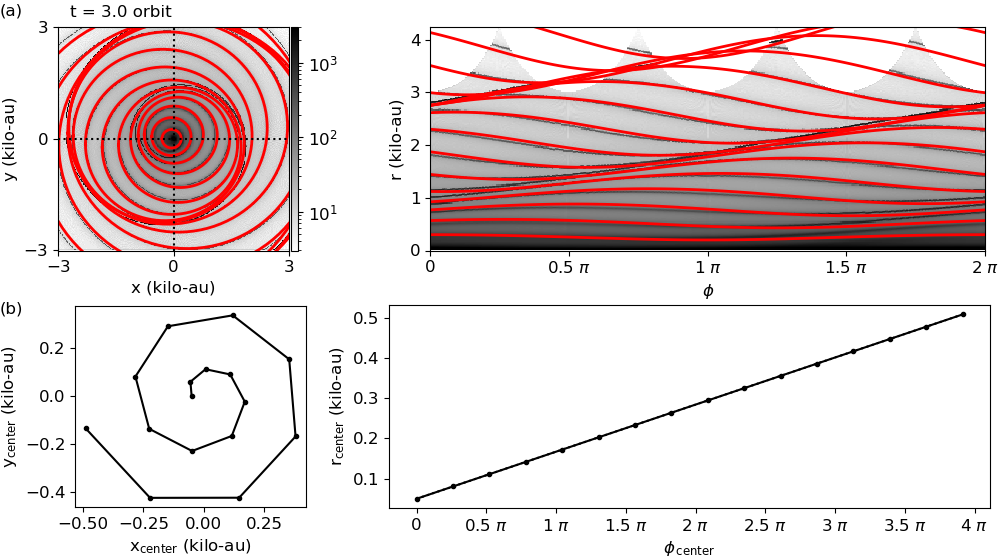}
  \caption{\label{fig:cir}
    (a) Ring approximation for the locations of the fine spirals (red),
    overlaid upon Figure\,\ref{fig:den}(d), the sticky particle model, in
    grayscale. (b) The positions of the centers of the rings illustrated
    in panel (a) with respect to the systemic center of mass. The increment
    of ring radius in each of the inner orbits is 0.26\,kilo-au.
  }
\end{figure*}

In Figure\,\ref{fig:cir}(a), the circular rings illustrated in red very well approximate the fine spirals.
These circular rings (or the components of the fine spiral) that has the time interval corresponding to the inner orbital period of $T_{\rm A-C}\sim85$\,yr converge on and form the main spiral pattern, which has a time interval corresponding to the outer orbital period of $T_{\rm AC-B}\sim690$\,yr. Therefore, the seemingly position-angle-dependent change in pattern interval is an illusion caused by the offsets of the centers of curvature.
Also, note that the centers of the rings follow a spiral that can be expressed by Equation\,(\ref{eqn:slp}) but for $T_{\rm A-C}$ instead of $T_{\rm AC-B}$ (see Figure\,\ref{fig:cir}(b)). The radii of the rings increase by a constant value of 260\,au during each of the inner orbits.

On the other hand, as shown in Figures\,\ref{fig:ecc}(c) and (d), in the models with the outer companion moving along a highly eccentric orbit, the fine spirals can be approximated by rings centered along the $x$-axis, which is the line connecting the pericenter and apocenter of the orbit.

\section{Summary and Discussion}\label{sec:dis}

Three-body systems are being recognized as important for explaining complex circumstellar structures and as a possible driver for the observed morphologies of a subset of objects undergoing the transition between the late stellar evolutionary phases from AGB to PN. In this paper, we have investigated the hydrodynamic and kinematic influences on the morphology of the expanding circumstellar envelope of a mass-losing star when it is being orbited not only by a relatively distant companion, but also by a third star that is closely orbiting the mass-losing star. We propose a triple system for the AGB star, CW Leo, which accounts for its complex circumstellar shell pattern, such as the off-centering of the ring-like pattern and the abrupt radial change in the interval of this pattern. 

We have first demonstrated the circumstellar pattern induced by a triple system by performing a hydrodynamic simulation in which the outflowing gas ejected by the AGB star is affected only by its initial velocity and the mass of the AGB star. We then compared the resulting density and velocity fields of the gas to the full three-body hydrodynamic simulation in which the gas response to the mass of all three bodies was computed, revealing incidental substructures originating from the gravitational wakes of companions.

The hydrodynamical effects are elucidated through comparison with the pinwheel model, which is nonhydrodynamic as simply following particles ejected isotropically from the mass-losing star having a predefined orbital motion. The pinwheel model mimics the circumstellar spiral pattern of the hydrodynamic model, exactly coinciding in location. However, it does not reproduce the density and width of the pattern, thereby highlighting the hydrodynamical effect upon the fluid as it encounters the shocks at the inner and outer edges of the dense ridge of the spiral pattern.

As a result, the density and density contrast of the fine spiral pattern (newly revealed to occur in a triple system) quickly decrease as the radius increases, while those along the main spiral pattern decline more slowly, just as in a two-body system. Therefore, the fine pattern may be observable only in the inner portions of the circumstellar envelope, making the observed image appear to have undergone an abrupt change in the pattern interval, as has been claimed for CW Leo \citep[e.g.,][]{gue18}.

Furthermore, the fine spiral pattern can be very well represented by off-centered circular rings having a regular radius increment. The center positions of the rings are located along a spiral, with the slope of that spiral in polar coordinates being related to the orbital period of the inner binary. The approximating off-centered rings coincide tangentially and therefore stack up along the ridge of the main spiral.

The detailed structure of the expansion velocity of the outflowing envelope is closely tied to the shape of the main spiral; after crossing the main spiral outward, the velocity quickly drops and slowly recovers its maximum value before reaching the next winding of the main spiral. Superimposed on that large-scale velocity pattern, the velocity jumps at the locations of the fine spiral pattern are much smaller. If the propagation speed of the matter is measurable, the detailed variations in the expansion velocity could contribute to the discrimination between the main and fine spirals in a triple system.

The dependence of the pattern on orbital eccentricities and inner companion mass has also been explored. A larger orbital eccentricity of the outer companion creates a density pattern that tightens a ``bowtie'' with the ``knot'' at the position angle toward the pericenter of the mass-losing star. And the density outside the bowtie significantly drops, causing a one-sided gap in the inter-pattern density. A larger orbital eccentricity of the inner companion broadens the width of the fine spiral, in particular at angles corresponding to the apocenter of the mass-losing star. Finally, a more massive inner companion reduces the density in the regions between the ridges of both the main and fine spiral patterns, thereby raising the density contrast of the pattern.

According to the results of our model calculations, CW Leo's abrupt radial change in the interval of the pattern and the off-centering of the pattern with intertwining ridges may hint at its nature as a triple system. As CW Leo has been getting brighter for already two decades, possibly indicating its evolutionary phase at the critical AGB--pPN transition moment \citep{kim21}, continuous monitoring of it would be beneficial to understanding the roles of the companions, in particular of the inner companion, in the morphological transformation. 
  
In the case of another example, the AGB star $\pi^1$ Gru, recent ALMA observations reveal an HCN spiral having an expansion time interval between successive turns of $\sim10$\,yr, which is consistent with the orbital period of a potential companion located at the secondary continuum peak, assuming that it has about one solar mass \citep{hom20}; the implied companion differs from the known G0V-type companion located 10 times further from the mass-losing star \citep{fea53,ake92}, plausibly suggesting a triple system with an orbital period ratio of $>500$. However, the CO spiral found by \citet{hom20} has an interval of $\sim80$\,yr, which is inconsistent with either of the above orbital timescales. It remains unclear whether this implies a fourth object in this system, or whether this is a triple system having a large eccentricity and/or noncoplanar orbits.

Only the midplane structures are examined in this paper. The three-dimensional structure of the pattern induced by a triple stellar system and the position-velocity diagnostics for interpretation of spectro-imaging observations are deferred to a future study.
In this paper, the orbits of three stellar objects are assumed to be coplanar for simplicity. It would be interesting to study the complexity of the spiral-shell patterns by relaxing this orbital configuration.
Dynamical stability consideration for the orbits will also need to be pursued in the future, especially for the eccentric orbit cases, since only a subset of triple systems will be stable on timescales long enough to produce a mass-losing AGB star. In a different subset of triple systems, the radius of the AGB star will ultimately swell to a size comparable to the radius of the inner orbit, creating a common-envelope binary, thereby leaving the original triple system in a well-separated binary configuration.

\acknowledgments
We are grateful to the anonymous referee for comments that helped to improve this paper, and to Dr.\ Smadar Naoz for her valuable comments on orbit stability in triple systems.
This research was supported by the National Research Foundation of Korea (NRF) grant (No.\ NRF-2021R1A2C1008928) and the Korea Astronomy and Space Science Institute (KASI) grant (Project No.\ 2023-1-840-00), both funded by the Korean government (MSIT).
J.H.\ acknowledges the support of NSFC project 11873086 and the sponsorship (in part) of the Chinese Academy of Sciences (CAS) through a grant to the CAS South America Center for Astronomy (CASSACA) in Santiago, Chile.

\begin{thebibliography}{}
\bibitem[Ake \& Johnson(1992)]{ake92} Ake, T.~B., Johnson, H.~R.\ 1992, AAS Meeting Abstract, 180, BAAS, 24, 788
\bibitem[Balick \& Frank(2002)]{bal02} Balick, B., Frank, A.\ 2002, \araa, 40, 439. doi:10.1146/annurev.astro.40.060401.093849  
\bibitem[Cernicharo et al.(2015)]{cer15} Cernicharo, J., Marcelino, N., Ag{\'u}ndez, M., Gu{\'e}lin, M.\ 2015, \aap, 575, A91. doi:10.1051/0004-6361/201424565
\bibitem[De Marco(2009)]{dem09} De Marco, O.\ 2009, \pasp, 121, 316. doi:10.1086/597765
\bibitem[Decin et al.(2020)]{dec20} Decin, L., Montarg{\`e}s, M., Richards, A.~M.~S., et al.\ 2020, Sci, 369, 1497. doi:10.1126/science.abb1229
\bibitem[Decin et al.(2015)]{dec15} Decin, L., Richards, A.~M.~S., Neufeld, D., et al.\ 2015, \aap, 574, A5. doi:10.1051/0004-6361/201424593
\bibitem[Duquennoy \& Mayor(1991)]{duq91} Duquennoy, A., Mayor, M.\ 1991, \aap, 248, 485
\bibitem[Feast(1953)]{fea53} Feast, M.~W.\ 1953, \mnras, 113, 510
\bibitem[Fryxell et al.(2000)]{fry00} Fryxell, B., Olson, K., Ricker, P., et al.\ 2000, \apjs, 131, 273. doi:10.1086/317361
\bibitem[Groenewegen et al.(2012)]{gro12} Groenewegen, M.~A.~T., Barlow, M.~J., Blommaert, J.~A.~D.~L., et al.\ 2012, \aap, 543, L8. doi:10.1051/0004-6361/201219604
\bibitem[Guelin et al.(1993)]{gue93} Guelin, M., Lucas, R., Cernicharo, J.\ 1993, \aap, 280, L19
\bibitem[Gu{\'e}lin et al.(2018)]{gue18} Gu{\'e}lin, M., Patel, N.~A., Bremer, M., et al.\ 2018, \aap, 610, A4. doi:10.1051/0004-6361/201731619  
\bibitem[He(2007)]{he07} He, J.~H.\ 2007, \aap, 467, 1081. doi:10.1051/0004-6361:20066435
\bibitem[Hirsch et al.(2021)]{hir21} Hirsch, L.~A., Rosenthal, L., Fulton, B.~J., et al.\ 2021, \aj, 161, 134. doi:10.3847/1538-3881/abd639
\bibitem[Homan et al.(2020)]{hom20} Homan, W., Montarg{\`e}s, M., Pimpanuwat, B., et al.\ 2020, \aap, 644, A61. doi:10.1051/0004-6361/202039185
\bibitem[Jeffers et al.(2014)]{jef14} Jeffers, S.~V., Min, M., Waters, L.~B.~F.~M., et al.\ 2014, \aap, 572, A3. doi:10.1051/0004-6361/201423463
\bibitem[Kim(2023)]{kim23} Kim, H.\ 2023, JKAS, 56, 149. doi:10.5303/JKAS.2023.56.2.149
\bibitem[Kim et al.(2013)]{kim13} Kim, H., Hsieh, I.-T., Liu, S.-Y., Taam, R.~E.\ 2013, \apj, 776, 86. doi:10.1088/0004-637X/776/2/86
\bibitem[Kim et al.(2021)]{kim21} Kim, H., Lee, H.-G., Ohyama, Y., et al.\ 2021, \apj, 914, 35. doi:10.3847/1538-4357/abf6cc
\bibitem[Kim et al.(2015)]{kim15} Kim, H., Liu, S.-Y., Hirano, N., et al.\ 2015, \apj, 814, 61. doi:10.1088/0004-637X/814/1/61
\bibitem[Kim et al.(2019)]{kim19} Kim, H., Liu, S.-Y., Taam, R.~E.\ 2019, \apjs, 243, 35. doi:10.3847/1538-4365/ab297e
\bibitem[Kim \& Taam(2012a)]{kim12a} Kim, H., Taam, R.~E.\ 2012a, \apj, 744, 136. doi:10.1088/0004-637X/744/2/136
\bibitem[Kim \& Taam(2012b)]{kim12b} Kim, H., Taam, R.~E.\ 2012b, \apj, 759, 59. doi:10.1088/0004-637X/759/1/59
\bibitem[Kim et al.(2017)]{kim17} Kim, H., Trejo, A., Liu, S.-Y., et al.\ 2017, NatAs, 1, 0060. doi:10.1038/s41550-017-0060
\bibitem[Kozai(1962)]{koz62} Kozai, Y.\ 1962, \aj, 67, 591. doi:10.1086/108790
\bibitem[Li et al.(2014)]{li14} Li, G., Naoz, S., Kocsis, B., et al.\ 2014, \apj, 785, 116. doi:10.1088/0004-637X/785/2/116
\bibitem[Lidov(1962)]{lid62} Lidov, M.~L.\ 1962, P\&SS, 9, 719. doi:10.1016/0032-0633(62)90129-0
\bibitem[Maercker et al.(2012)]{mae12} Maercker, M., Mohamed, S., Vlemmings, W.~H.~T., et al.\ 2012, Natur, 490, 232. doi:10.1038/nature11511
\bibitem[Mastrodemos \& Morris(1999)]{mas99} Mastrodemos, N., Morris, M.\ 1999, \apj, 523, 357. doi:10.1086/307717
\bibitem[Mauron \& Huggins(1999)]{mau99} Mauron, N., Huggins, P.~J.\ 1999, \aap, 349, 203
\bibitem[Men'shchikov et al.(2001)]{men01} Men'shchikov, A.~B., Balega, Y., Bl{\"o}cker, T., et al.\ 2001, \aap, 368, 497. doi:10.1051/0004-6361:20000554
\bibitem[Parker(1958)]{par58} Parker, E.~N.\ 1958, \apj, 128, 664. doi:10.1086/146579
\bibitem[Raghavan et al.(2010)]{rag10} Raghavan, D., McAlister, H.~A., Henry, T.~J., et al.\ 2010, \apjs, 190, 1. doi:10.1088/0067-0049/190/1/1
\bibitem[Ramos-Larios et al.(2016)]{ram16} Ramos-Larios, G., Santamar{\'\i}a, E., Guerrero, M.~A., et al.\ 2016, \mnras, 462, 610. doi:10.1093/mnras/stw1572
\bibitem[Sahai et al.(2007)]{sah07} Sahai, R., Morris, M., S{\'a}nchez Contreras, C., et al.\ 2007, \aj, 134, 2200. doi:10.1086/522944
\bibitem[Sahai et al.(2011)]{sah11} Sahai, R., Morris, M.~R., Villar, G.~G.\ 2011, \aj, 141, 134. doi:10.1088/0004-6256/141/4/134
\bibitem[Sahai et al.(2016)]{sah16} Sahai, R., Scibelli, S., Morris, M.~R.\ 2016, \apj, 827, 92. doi:10.3847/0004-637X/827/2/92
\bibitem[Salas et al.(2019)]{sal19} Salas, J.~M., Naoz, S., Morris, M.~R., et al.\ 2019, \mnras, 487, 3029. doi:10.1093/mnras/stz1515
\bibitem[Shivamoggi et al.(2021)]{shi21} Shivamoggi, B., Rollins, D., Pohl, L.\ 2021, Entrp, 23, 1497. doi:10.3390/e23111497
\bibitem[Soker(1994)]{sok94} Soker, N.\ 1994, \mnras, 270, 774. doi:10.1093/mnras/270.4.774
\bibitem[Theuns \& Jorissen(1993)]{the93} Theuns, T., Jorissen, A.\ 1993, \mnras, 265, 946. doi:10.1093/mnras/265.4.946
\bibitem[Tokovinin(2001)]{tok01} Tokovinin, A.\ 2001, in IAU Symp.\ 200, ASP Conf.~Ser.\ The Formation of Binary Stars, ed.\ H.~Zinnecker \& R.~D.~Mathieu (Cambridge: Cambridge Univ.\ Press), 84
\bibitem[Tokovinin(2021)]{tok21} Tokovinin, A.\ 2021, Univ, 7, 352. doi:10.3390/universe7090352
\bibitem[Tokovinin et al.(2006)]{tok06} Tokovinin, A., Thomas, S., Sterzik, M., et al.\ 2006, \aap, 450, 681. doi:10.1051/0004-6361:20054427
\bibitem[Truelove \& McKee(1999)]{tru99} Truelove, J.~K., McKee, C.~F.\ 1999, \apjs, 120, 299. doi:10.1086/313176
\bibitem[Zuckerman \& Aller(1986)]{zuc86} Zuckerman, B., Aller, L.~H.\ 1986, \apj, 301, 772. doi:10.1086/163943
\end{thebibliography}
\end{document}